\documentclass[a4paper,twoside,10pt]{letter}
\usepackage{graphicx,saj,multicol,subeqnarray}
\def\papertype{Original Scientific Paper}
\def\degr{\hbox{$^\circ$}}
\def\arcmin{\hbox{$^\prime$}}
\def\arcsec{\hbox{$^{\prime\prime}$}}
\def\SNR{\mbox{{SNR~J0519--6926}}}

\begin{document}
\baselineskip=3.1truemm
\columnsep=.5truecm
\newenvironment{lefteqnarray}{\arraycolsep=0pt\begin{eqnarray}}
{\end{eqnarray}\protect\aftergroup\ignorespaces}
\newenvironment{lefteqnarray*}{\arraycolsep=0pt\begin{eqnarray*}}
{\end{eqnarray*}\protect\aftergroup\ignorespaces}
\newenvironment{leftsubeqnarray}{\arraycolsep=0pt\begin{subeqnarray}}
{\end{subeqnarray}\protect\aftergroup\ignorespaces}
%

% Running titles

\markboth{\eightrm RADIO CONTINUUM STUDY OF SNR's IN THE LMC -- \SNR}
{\eightrm E.J.~Crawford, M.D.~Filipovi\'c and  J.L.~Payne}

{\ }

\publ

\type

{\ }

% Title
\title{RADIO CONTINUUM STUDY OF SUPERNOVA REMNANTS IN THE LARGE MAGELLANIC CLOUD -- \SNR}

% Authors

\authors{E.J.~Crawford$^{1}$, M.D.~Filipovi\'c$^1$ and J.L.~Payne$^2$}

\vskip3mm

% Address

\address{$^1$School of Computing and Mathematics,
University of Western Sydney\break Locked Bag 1797, Penrith South DC, NSW 1797, Australia}

\address{$^2$Centre for Astronomy, James Cook University
\break Townsville, QLD 4811, Australia}

% Received and Accepted dates

\dates{February, 2008}{February, 2008}

% Abstract

\summary{We present the results of new high resolution ATCA observations of \SNR. We found that this SNR exhibits a typical ``horseshoe'' appearance with $\alpha$=--0.55$\pm$0.08 and D=28$\pm1$~pc. No polarization (or magnetic fields) are detected to a level of 1\%. This is probably due to a relatively poor sampling of the $uv$ plane caused be observing in ``snap-shot'' mode. }

% Keywords (see keywords.pdf file)

\keywords{ISM: supernova remnants -- Magellanic Clouds -- Radio Continuum: ISM --  SNR B0520--69}

\begin{multicols}{2}
{
% Sections

\section{1. INTRODUCTION}

The study of radio supernova remnants (SNRs) in nearby galaxies is of major interest in order to understand the radio output of more distant galaxies and to understand the processes that occur on local interstellar scales within our own Galaxy. Unfortunately, the distances to many Galactic remnants are uncertain by a factor of 2, leading to a factor of 4 uncertainty in luminosity and of 5.5 in the calculated energy release of the initiating supernova (SN). At a distance of $\sim$50~kpc (Hilditch et al. 2005), the Large Magellanic Cloud (LMC) is one of the prime targets for the astrophysical research of galactic objects, including SNRs. This is because these remnants are located at a known distance, yet close enough to allow a detailed analysis of them.

SNRs reflect a major process in the elemental enrichment of the interstellar medium (ISM). Multiple supernova explosions close together generate super-bubbles typically hundreds of parsecs in extent. Both are among the prime drivers controlling the morphology and the evolution of the ISM. Their properties are therefore crucial for the full understanding of the galactic matter cycle.

Radio observations are commonly used to discover and characterize SNRs, as the non-thermal emission 
typical of SNR's is easily detectable at centimeter  radio wavelengths. Verification may be based on a combination of radio spectral index, morphology, co-identifications in other domains (such as optical and/or \mbox{X-ray}), flux density and location within the galaxy under study. There are over 40 confirmed SNRs in the LMC and another 35-40 candidates (Payne et al. 2008). 

\SNR\ was initially classified as an SNR based on the Einstein X-Ray survey by Long et al.\ (1981) 
(named LHG~27) and Wang et al. (1991) (named W~19). Mathewson et al. (1983) catalogued \SNR\ as B0520-69.4
based on their optical observations, reporting an estimated size of $138\arcsec\times104\arcsec$. Filipovi\'c et al. (1998) added further confirmation, with a set of radio continuum observations (with the Parkes telescope) on a wide frequency range. Blair et al. (2006) report detection at far ultraviolet (FUV) wavelengths. Haberl \& Pietsch (1999) (named SNR as HP~915) and Williams et al. (1999) discuss the X-Ray properties of \SNR\ based on ROSAT observations. Most recently Payne et al. (2008) presented optical spectroscopy of a wide range of LMC SNRs including \SNR. They found an enhanced [S\textsc{ii}]/H$_\alpha$ ratio of 0.8 typical for SNRs.

\section{2. OBSERVATION DATA}

We observed \SNR\ with the Australia Telescope Compact Array (ATCA) on 6$^\mathrm{th}$ April 1997, with array configuration 375-m, at wavelengths of 6 and 3~cm ($\nu$=4790 and 8640~MHz). The observations were done in so called ``snap-shot'' mode, totaling $\sim$1 hour of integration over a 12 hour period. Source 1934-638 was used for primary calibration and source 0530-727 was used for secondary calibration. The \textsc{miriad} (Sault \& Killeen 2006) and \textsc{karma} (Gooch 2006) software packages were used for reduction and analysis. More information about observing procedure and another source (LMC SNR B0513-69) 
observed during this session can be found in Boji\v ci\'c et al. (2007).

Images were prepared, cleaned and deconvolved using \textsc{miriad} tasks. Baselines formed with the $6^\mathrm{th}$ ATCA antenna were excluded, as the other five antennas were arranged in a compact configuration. The 6~cm image (Fig.~1) has a resolution of 32\arcsec\ and the r.m.s noise is estimated to be 0.4~mJy/beam. Similarly, the 3~cm image (Fig.~2) has a resolution of 20\arcsec\ and the r.m.s noise is estimated to be 0.4~mJy/beam. A 3~cm image was also created with a resolution of 32\arcsec\ to compliment the 6~cm image, and this was used in the preparation of Figs.~1~and~3.

\section{3. RESULTS AND DISCUSSION}

The remnant has a classical horseshoe morphology centered at RA(J2000)=5$^h$19$^m$45.3$^s$, DEC(J2000)=--69\degr26\arcmin 00.6\arcsec\ with a nearly circular diameter of 140\arcsec$\pm$5\arcsec\ (28$\pm$1~pc). This is reasonably consistent with the optical size reported by Mathewson et al. (1983), and in agreement with the X-ray size $156\arcsec\times132\arcsec$ reported by Williams et al. (1999).

Flux density measurements were made at 6 and 3~cm, resulting in values of 71.3~mJy and 43.5~mJy respectively (Table~1). We made our own measurement of flux density at 1400~MHz using the mosaic image described in Hughes et al. (2007). A spectral index (defined as $S\propto\nu^\alpha$) was plotted using the flux densities in Table~1 (Fig.~4) and estimated to be $\alpha$=--0.55$\pm$0.08, which is typical for SNRs (Filipovi\'c et al. 1998). We note that there are two points (843 and 2300~MHz) which lie slightly off the line of best fit. The estimate of flux density for these points are very uncertain, and the error bars are probably too conservative. This is  due to missing short spacings effects at 843~MHz. At 2300~MHz a large r.m.s noise coupled with a large Parkes beam size accounts for the errors. We also calculated a two point spectral index between 6 and 3~cm and found $\alpha$=--0.84. This is somewhat steeper than the overall $\alpha$=--0.55 but also expected as for most SNRs the flux density is fading with increasing frequency.

From our spectral index map (Fig.~3) we can see that spectral index is changing rapidly at the SNR front shock with the usual $\alpha$$\approx$--0.8, due to the non-thermal nature of emission in the shocked regions. Further on, inside the SNR shell, $\alpha$ tends to be flatter ($-0.4$$\le$$\alpha$$\le$$0$) as the emission has larger  thermal component. We note that there are two ``holes'' in the spectral index map (Fig.~3) were thermal emission ($\alpha$$\ge$0) from this SNR tends to dominate. Comparing our radio-continuum images with ROSAT HRI X-ray images of Williams et al. (1999; Fig.~1l) shows an anti-correlation between radio and X-ray emission.

Linear polarization images for each frequency were created using \textit{Q} and \textit{U} parameters. While we detect no reliable polarization at 8640~MHz, the 4790~MHz image reveals some very low level linear
polarization. Without reliable polarization measurements at the second frequency we could not
determine if any Faraday rotation was present.

The mean fractional polarization at 4790~MHz was calculated using flux density and polarization:
\begin{equation}
P=\frac{\sqrt{S_{Q}^{2}+S_{U}^{2}}}{S_{I}}\cdot 100\%
\end{equation}
\noindent where $S_{Q}, S_{U}$ and $S_{I}$ are integrated
intensities for \textit{Q}, \textit{U} and \textit{I} Stokes
parameters. Our estimated value is $P\cong 1\%$.

\section{4. CONCULUSION}

We conducted the highest resolution observations to date of \SNR. From these observations be found a diameter of $140\arcsec\pm5\arcsec$, a spectral index $\alpha$=--0.55$\pm$0.08 and no linear polarization to a level of $\sim1\%$.
}
\end{multicols}
\vfil\clearpage
\centerline{{\bf Table 1.} Integrated Flux.}
\vskip2mm
\centerline{\begin{tabular}{|c|c|c|c|c|c|c|}
\hline
 &S$_\mathrm{I}$ (408~MHz)&S$_\mathrm{I}$ (843~MHz)&S$_\mathrm{I}$ (1400~MHz)&S$_\mathrm{I}$ (2300~MHz)&S$_\mathrm{I}$ (4790~MHz)&S$_\mathrm{I}$ (8640~MHz) \\
 \hline
 \SNR          &280 mJy&141 mJy&135 mJy&124 mJy&71.3 mJy&43.5 mJy \\
 \hline
 Reference&Clarke        &Mills          & This               &Filipovi\'c    &This  &This \\
                    &et al. 1976 &et al. 1984& Work &et al. 1996&Work&Work \\
 \hline
\end{tabular}}
\vspace{0.5cm}
\centerline{\includegraphics[width=.47\textwidth,angle=-90]{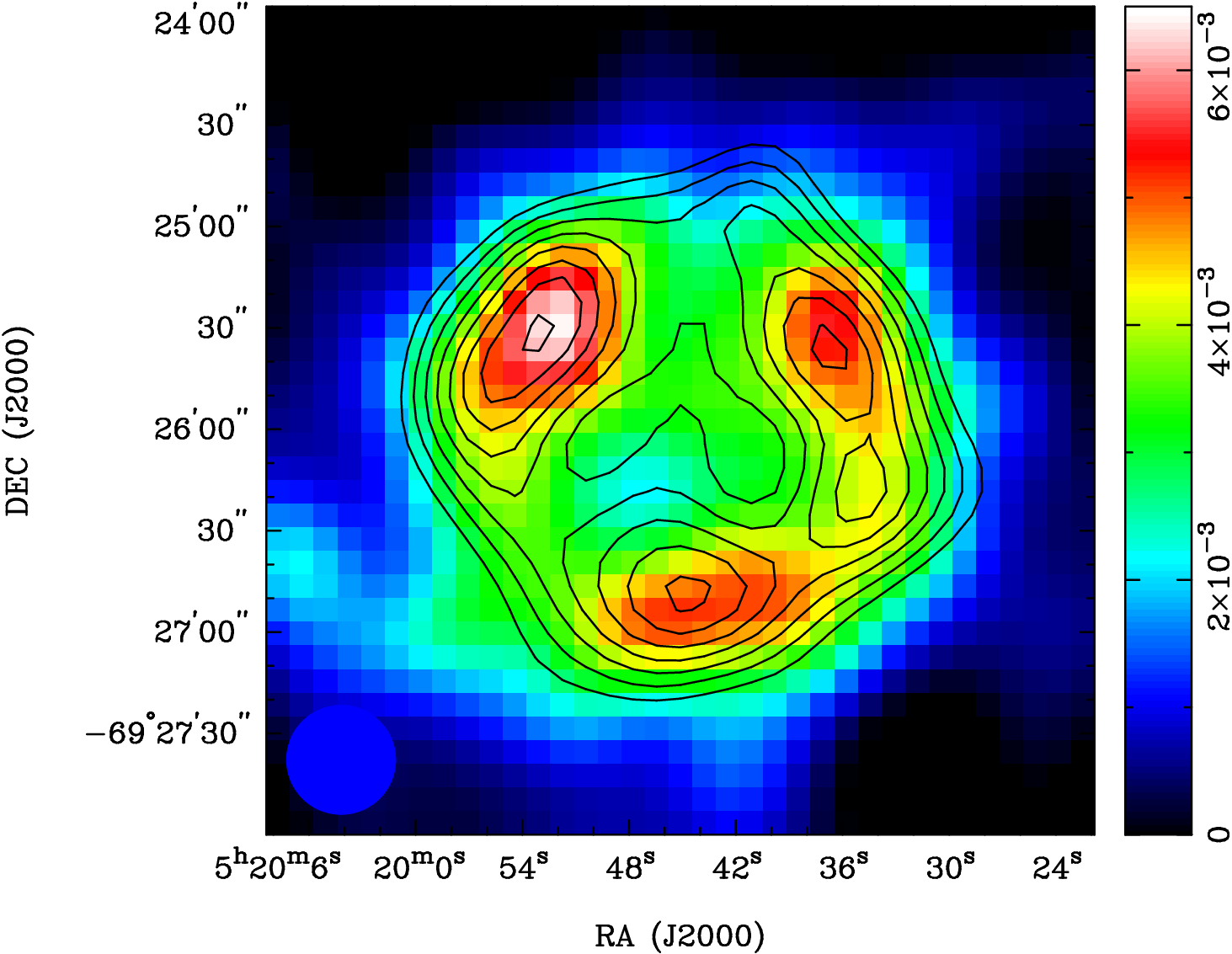}}
\figurecaption{1.}{ATCA observations of \SNR\ at 6~cm (4790~MHz) overlaid with 3~cm (8640~MHz) complimentary resolution contours. The contours are from 1.2 to 4~mJy/beam in 0.4~mJy/beam steps. The blue circle in the lower left corner represents the primary beam of 32\,\arcsec. The sidebar quantifies the pixel map and it's units are mJy/beam.}
\centerline{\includegraphics[width=.47\textwidth,angle=-90]{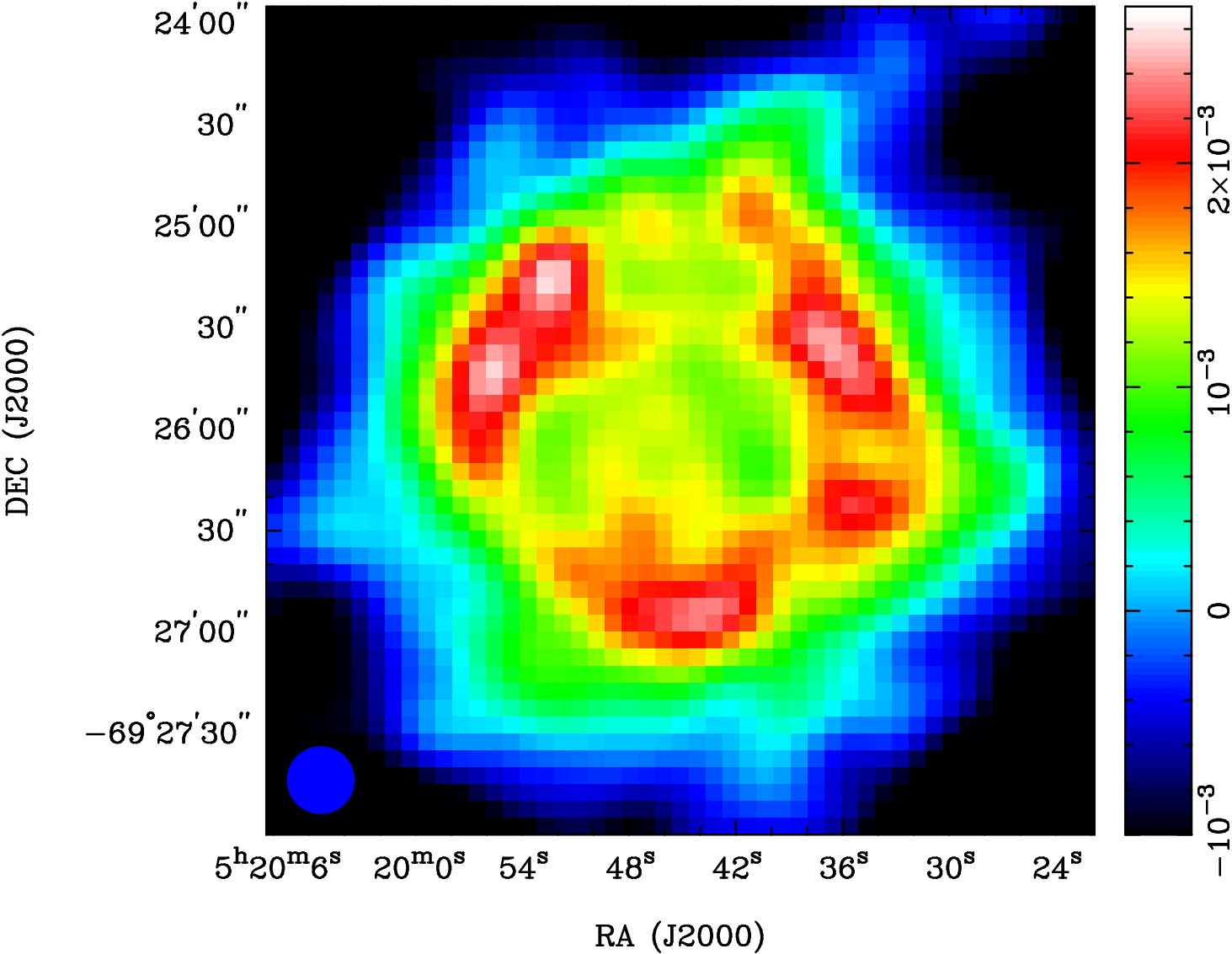}}
\figurecaption{2.}{ATCA 3cm observations of \SNR. The blue circle in the lower left corner represents the 3~cm (8640~MHz) primary beam of 20\,\arcsec. The sidebar quantifies the pixel map and it's units are mJy/beam.}
\centerline{\includegraphics[width=.47\textwidth,angle=-90]{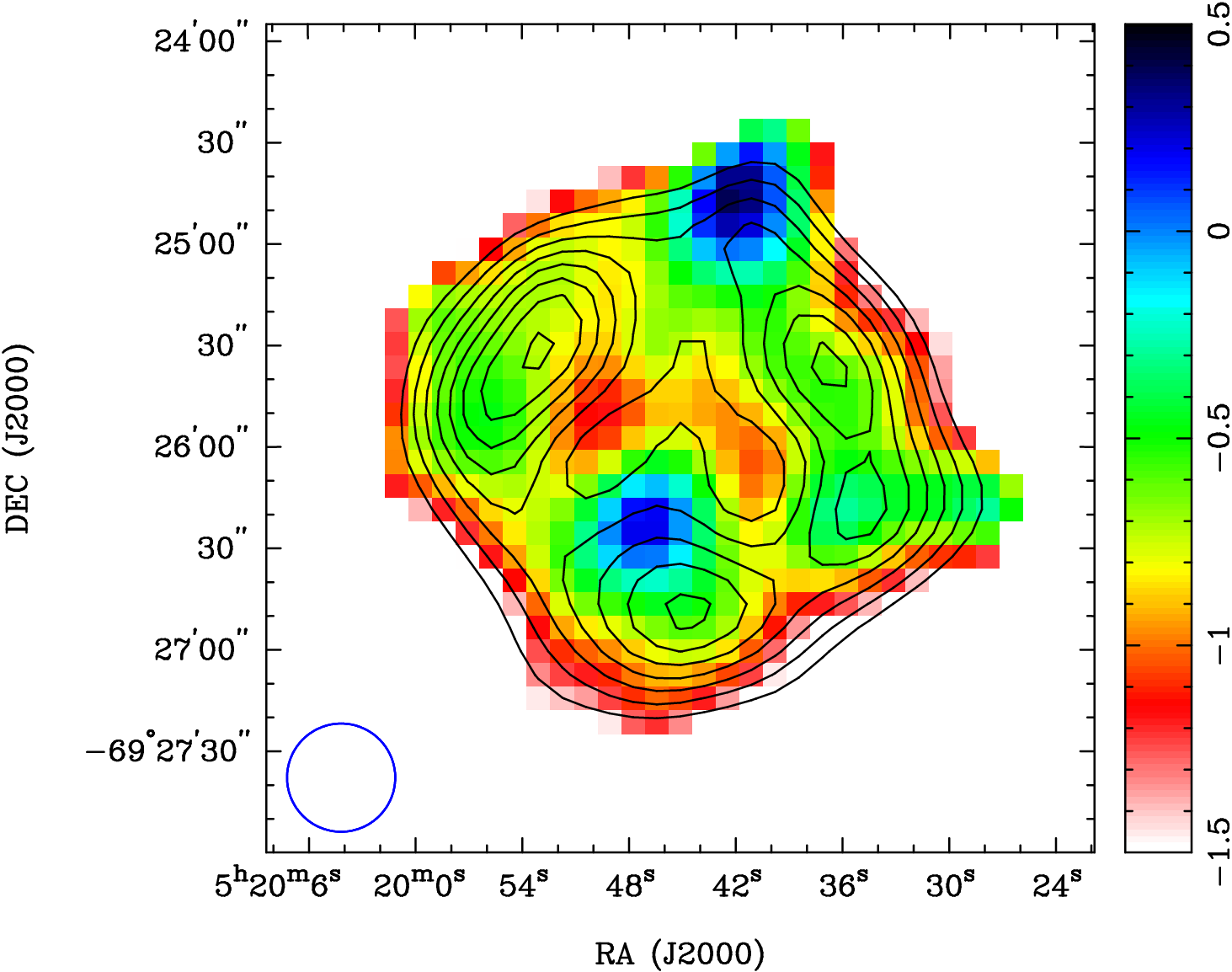}}
% Figure captions {number}{caption}
\figurecaption{3.}{Spectral index map of \SNR\ overlaid with 3~cm (8640~MHz) contours. The contours are from 1.2 to 4~mJy/beam in 0.4~mJy/beam steps. The blue circle in the lower left corner represents the primary beam of 32\,\arcsec. The sidebar quantifies the pixel map, representing the spectral index.}
\centerline{\includegraphics[width=0.47\textwidth,angle=-90]{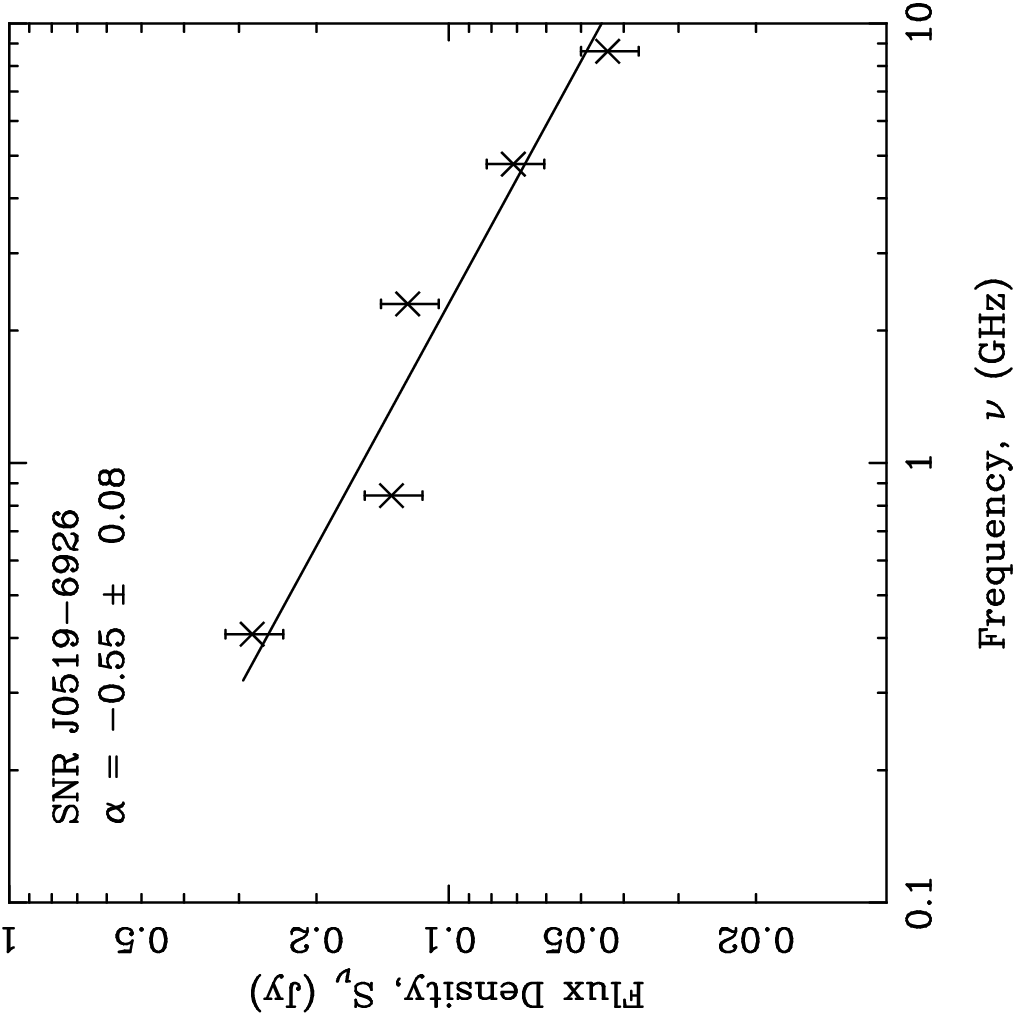}}
\figurecaption{4.}{Radio Spectrum of \SNR}

\vfil\clearpage

\begin{multicols}{2}
{
% Acknowledgements

\acknowledgements{We used the {\sc karma} software package developed by the ATNF. The Australia Telescope Compact Array is part of the Australia Telescope which is funded by the Commonwealth of Australia for operation as a National Facility managed by CSIRO.}

% References
\references
Blair, W.~P., Ghavamian, P., Sankrit, R., Danforth, C.~W.: 2006, \journal{Astron. Astrophys. Suppl. Ser.}, \vol{165}, 480

Boji{\v c}i{\'c}, I.~S., Filipovi{\'c}, M.~D., Parker, Q.~A., Payne, J.~L., Jones, P.~A., Reid, W., Kawamura, A.,  Fukui, Y.\ 2007, \journal{Mon. Not. R. Astron. Soc.}, \vol{378}, 1237 

Clarke, J.~N., Little,  A.~G.,  Mills, B.~Y.: 1976, \journal{Aust. J. Phys. Astrophys. Suppl.}, \vol{40}, 1

Filipovi\'c, M.~D., White, G.~L.,  Haynes, R.~F., Jones, P.~A., Meinert, D., Wielebinski, R.,  Klein, U.: 1996, \journal{Astron. Astrophys. Suppl. Ser.}, \vol{120}, 77

Filipovi\'c, M.~D., Haynes, R.~F., White, G.~L.,  Jones, P.~A.: 1998, \journal{Astron. Astrophys. Suppl. Ser.}, \vol{130}, 421

Gooch, R.: 2006, Karma Users Manual, ATNF

Haberl, F., Pietsch, W.: 1999, \journal{Astron. Astrophys. Suppl. Ser.}, \vol{139}, 277 
      
Hilditch, R.~W., Howarth, I.~D., Harries, T.~J.: 2005, \journal{Mon. Not. R. Astron. Soc.}, \vol{ 357}, 304

Hughes, A., Staveley-Smith, L., Kim, S., Wolleben, M.,  Filipovi{\'c}, M.\ 2007,  \journal{Mon. Not. R. Astron. Soc.}, \vol{382}, 543
 
Long, K.~S., Helfand, D.~J.,  Grabelski, D.~A.: 1981, \journal{ApJ}, \vol{248}, 925

Mathewson, D.~S., Ford, V.~L., Dopita, M.~A., Tuohy, I.~R., Long, K. ~S., Helfand, D.~J.:1983, \journal{Astron. Astrophys. Suppl. Ser.}, \vol{51}, 345

Mills, B.~Y., Turtle, A.~J., Little, A.~G.,  Durdin, J.~M.: 1984, \journal{Aust. J. Phys.}, \vol{37}, 321

Payne, J.~L., White, G.~L., Filipovi{\'c}, M.~D.\ 2008, \journal{Mon. Not. R. Astron. Soc.}, \vol{383}, 1175

Sault, R., Killeen, N.: 2006, Miriad Users Guide, ATNF

Wang, Q., Hamilton, T., Helfand, D.~J., Wu X.: 1991, \journal{ApJ}, \vol{374}, 475

Williams, R.~M., Chu, Y.-H., Dickel, J.~R., Petre, R., Smith, R.~C.,  Tavarez, M.: 1999, \journal{Astron. Astrophys. Suppl. Ser.}, \vol{123}, 467

\endreferences
}
\end{multicols}

\vfill\eject

{\ }

% Serbian abstract

% Title

\title{RADIO KONTINUM STUDIJA OSTATKA SUPERNOVE U VELIKOM MAGELANOVOM OBLAKU -- \SNR}

% Authors

\authors{E.J.~Crawford$^{1}$, M.D~Filipovi\'c$^1$ and J.L.~Payne$^2$}

\vskip3mm

% Address

\address{$^1$School of Computing and Mathematics,
University of Western Sydney\break Locked Bag 1797, Penrith South DC, NSW 1797, Australia}

\address{$^2$Centre for Astronomy, James Cook University
\break Townsville, QLD 4811, Australia}

\vskip.7cm

% UDC

%\centerline{UDK \udc}

% Papertype

\centerline{\rit }

\vskip.7cm

\begin{multicols}{2}
{

% Abstract

\rrm U ovoj studiji predstav{lj}amo nove ATCA radio-kontinum rezultate visoke rezulucije posmatra{nj}a ostatka supernove (OSN) u Velikom Magelanovom Oblaku -- \textrm{\SNR}. Naxli smo da ovaj OSN ima tipiqnu morfologiju (potkovica) za ovu vrstu objekata sa radio spektralnim indeksom od \mbox{$\alpha=-0.55\pm$0.08} i dijametrom od \mbox{D$=28\pm1$} parseka. Nismo detektovali polarizaciju (ni magnetno po{lj}e) do nivoa od 1\%. Razlog za ne postoja{nj}e polarizacije u naxim po\-sma\-tra\-{nj}ima treba tra{\zz}iti u tehnici ovih posmatra{nj}a koja su obav{lj}ena u tzv. ``snap-shot'' modu koji je pokrio samo ma{nj}i deo $uv$ po{lj}a. 

}
\end{multicols}

\end{document}